\documentclass[sigconf, nonacm]{acmart}

\usepackage{todonotes}
\usepackage{siunitx}
\usepackage[vlined, linesnumbered, ruled]{algorithm2e}
\usepackage{multirow}
\usepackage[subtle]{savetrees}

\graphicspath{{figures/}}

\SetCommentSty{commentfont}

\def\shine{\mbox{\textsc{Shine}}}
\def\shinebaseline{\mbox{\textsc{\shine-Baseline}}}
\def\shinecached{\mbox{\textsc{\shine-Cached}}}
\def\shinecachedaqr{\mbox{\textsc{\shine-Cached-Aqr}}}

\newcommand{\unitgb}[1]{\qty{#1}{\giga\byte}}
\newcommand{\unitgbit}[1]{\qty{#1}{\giga\bit}}
\newcommand{\unitmb}[1]{\qty{#1}{\mega\byte}}
\newcommand{\unitkb}[1]{\qty{#1}{\kilo\byte}}
\newcommand{\unitbyte}[1]{\qty{#1}{\byte}}
\newcommand{\unitbit}[1]{\qty{#1}{\bit}}
\newcommand{\unitbitadj}[1]{\SI[number-unit-product=\text{-}]{#1}{\bit}}

\def\rdmaread{\texttt{READ}}

\def\rdmacas{\texttt{CAS}}
\def\rdmacaslong{\texttt{RDMA\_CAS}}
\def\rdmafaa{\texttt{FAA}}
\def\rdmawrite{\texttt{WRITE}}
\def\rdmawritelong{\texttt{RDMA\_WRITE}}
\def\rdmasend{\texttt{SEND}}
\def\rdmareceive{\texttt{RECV}}

\def\bigann{\mbox{BIGANN}}
\def\deep{\mbox{DEEP}}
\def\spacev{\mbox{SPACEV}}
\def\tti{\mbox{TTI}}
\def\turing{\mbox{TURING}}

\newcommand{\regularinlinenode}[1]{%
\begin{tikzpicture}[baseline]%
  \node[font=\normalfont,draw,line width=0.5pt,rounded corners=0.5mm,text=black,fill=black!2,inner sep=1.75pt,anchor=base,yshift=0.8pt] {\scriptsize #1};%
\end{tikzpicture}%
}

\newcommand{\challenge}[2]{\regularinlinenode{C#1}\;\emph{#2}.\,}
\newcommand{\smallchallenge}[1]{\regularinlinenode{C#1}}

\newcommand\vldbdoi{XX.XX/XXX.XX}

\newcommand\vldbavailabilityurl{}
\newcommand\vldbpagestyle{plain} 

\begin{document}
\title{SHINE: A Scalable HNSW Index in Disaggregated Memory}

\author{Manuel Widmoser}
\affiliation{%
  \institution{University of Salzburg, Austria}
}
\email{manuel.widmoser@plus.ac.at}
\orcid{0009-0002-0239-1474}

\author{Daniel Kocher}
\affiliation{%
  \institution{University of Salzburg, Austria}
}
\email{dkocher@cs.sbg.ac.at}
\orcid{0009-0003-3742-5555}

\author{Nikolaus Augsten}
\affiliation{%
  \institution{University of Salzburg, Austria}
}
\email{nikolaus.augsten@plus.ac.at}
\orcid{0000-0002-3036-6201}

\begin{abstract}
  Approximate nearest neighbor (ANN) search is a fundamental problem in computer science for which in-memory graph-based methods, such as Hierarchical Navigable Small World (HNSW), perform exceptionally well.
To scale beyond billions of high-dimensional vectors, the index must be distributed.
The disaggregated memory architecture physically separates compute and memory into two distinct hardware units and has become popular in modern data centers. 
Both units are connected via RDMA networks that allow compute nodes to directly access remote memory and perform all the computations, posing unique challenges for disaggregated indexes.

In this work, we propose a scalable HNSW index for ANN search in disaggregated memory.
In contrast to existing distributed approaches, which partition the graph at the cost of accuracy, our method builds a graph-preserving index that reaches the same accuracy as a single-machine HNSW.
Continuously fetching high-dimensional vector data from remote memory leads to severe network bandwidth limitations, which we overcome by employing an efficient caching mechanism.
Since answering a single query involves processing numerous unique graph nodes, caching alone is not sufficient to achieve high scalability.
We logically combine the caches of the compute nodes to increase the overall cache effectiveness and confirm the efficiency and scalability of our method in our evaluation.

\end{abstract}

\maketitle

\pagestyle{\vldbpagestyle}

\ifdefempty{\vldbavailabilityurl}{}{
\vspace{.3cm}
\begingroup\small\noindent\raggedright\textbf{PVLDB Artifact Availability:}\\
The source code, data, and/or other artifacts have been made available at \url{\vldbavailabilityurl}.
\endgroup
}

\section{Introduction}

Vector nearest neighbor search aims to find similar vectors from a large data collection and plays a crucial role in various real-world applications, such as information retrieval~\cite{liu-07,zhang-22}, recommendation systems~\cite{schafer-07,chen-22}, visual search~\cite{zhang-18}, and retrieval-augmented generation~\cite{asai-23,gao-23}.
In $k$-nearest neighbor ($k$-NN) search, the goal is to find the $k$ most similar vectors with respect to a given query vector and a predefined distance function that assesses the similarity of two vectors, e.g., the Euclidean distance or the inner product.
However, computing the exact $k$-nearest neighbors in a high-dimensional vector space involves a linear scan over the entire dataset due to the curse of dimensionality~\cite{weber-98,indyk-98}, which is infeasible for large vector collections.
To overcome this problem, recent solutions focus on \emph{approximate} nearest neighbor (ANN) search and trade accuracy for performance.

Four categories of techniques to answer ANN queries have been explored:
(1) Tree-based methods~\cite{fukunaga-75,ram-12,muja-09},
(2) quantization-based methods~\cite{ge-13,gao-24,jegou-11,wu-17},
(3) LSH-based methods~\cite{gionis-99,indyk-98,shrivastava-12}, and
(4) graph-based methods~\cite{malkov-14,malkov-20,fu-19,hajebi-11,wu-14,morozov-18,wang-24}.
Recent empirical studies~\cite{li-20,wang-xu-21,aumuller-20} show that in-memory graph-based methods like Hierarchical Navigable Small World (HNSW)~\cite{malkov-20} perform exceptionally well for high-dimensional data.
Consequently, HNSW is one of the primary index structures in many vector database systems~\cite{pan-24}.
For very large vector collections, however, the non-negligible memory footprint of HNSW in combination with the high dimensionality of the vector data may exceed the main memory capacity of a single machine. 
For instance, storing the raw vector data (without the index structure) for one billion 200-dimensional floating-point vectors requires \unitgb{800} of main memory.
To scale horizontally to billions of vectors, distributed methods~\cite{doshi-21,deng-19,muja-14,liu-25} partition the vector space and build sub-indexes on smaller data subsets per machine~\cite{gottesburen-24,dong-23}. 
However, partitioning the graph requires removing potentially important edges, resulting in a trade-off between accuracy and efficiency.

Modern data centers employ a disaggregated memory architecture and decouple computation and memory into two separate hardware units that are connected via an ultra-fast low-latency network.
This opens up a new design space for graph-based index structures to solve the scalability problem of vector similarity search without partitioning the vector space.
Compute nodes have many CPU cores but only a few GBs of main memory, whereas memory nodes have near-zero computation power but large amounts of main memory.
With remote direct memory access (RDMA), a modern networking mechanism, a compute node is able to directly access the memory of a memory node without involving its CPU.
The architectural shift from monolithic servers to disaggregated memory allows to scale compute and memory independently, thereby optimizing the resource utilization.
For this reason, the disaggregated memory architecture has become prevalent in industry~\cite{pinto-20,keeton-15,cao-21,aguilera-23} and academia~\cite{guo-22,lee-21,shan-19,zhang-20}.

In this work, we present \shine, the first graph-preserving HNSW index for disaggregated memory.
Compared to other distributed approaches, \shine\ builds a \emph{global} HNSW index that retains all edges of a conventional HNSW graph. 
Therefore, \shine\ reaches the same accuracy as a single-machine HNSW index, although the index is distributed to arbitrarily many memory nodes.
In disaggregated memory, compute nodes have only limited memory resources and must fetch all the data needed for computation from the memory nodes.
Since each query must evaluate thousands of high-dimensional vectors, reading from remote memory forms a severe bottleneck.
To consume less network bandwidth, \shine\ leverages the limited memory on the compute nodes to cache frequently accessed vectors.
However, HNSW imposes particularly high demands on caching, due to the need to (i) fetch many distinct nodes and (ii) maintain high-dimensional vectors for distance computations.
Since the caches of the compute nodes are independent, each compute node must store many vectors already cached by other compute nodes, such that the aggregated cache size over all compute nodes is used ineffectively.
Therefore, we propose to logically combine the caches across the compute nodes to significantly increase the overall cache effectiveness.
We logically partition the HNSW graph and assign each partition to a dedicated compute node.
Whenever a compute node receives a query from a client, it routes the query to the node responsible for the best-matching partition. 
Routing is a lightweight task for which we exploit the near-zero computation power of the memory nodes and argue that this is not a bottleneck.
To avoid overloading individual compute nodes for skewed query workloads, \shine\ tracks their progress and adaptively adjusts the query routes to balance the workload.

We implement \shine\ and evaluate its performance and scalability on various large-scale datasets.  
To the best of our knowledge, \shine\ is the first distributed HNSW index specifically designed for disaggregated memory and the first distributed approach that retains the accuracy of a single-machine HNSW graph.

\paragraph{Contributions}
Our contributions can be summarized as follows:
\newline\indent $\bullet$\,
We propose \shine, the first graph-preserving HNSW index design under memory disaggregation.
In contrast to existing distributed solutions, our method does \emph{not} break the internal graph structure and retains the accuracy of a conventional HNSW index.
\newline\indent $\bullet$\,
We leverage the low memory on compute nodes and the low compute power on memory nodes to significantly boost the performance of graph-based indexes.
In particular, we integrate compute-side caching with \emph{logical index partitioning} and \emph{adaptive query routing} to improve the overall cache effectiveness.
\newline\indent $\bullet$\,
We formally introduce a \emph{cache segmentation penalty} to quantify the effect of segmenting the cache across compute nodes  w.r.t.\ an (hypothetical) single large cache accessible to all compute nodes.
\newline\indent $\bullet$\,
In our extensive empirical evaluation, we demonstrate the efficiency and scalability of \shine\ on various large-scale vector datasets.
Our evaluation confirms that the combination of our proposed optimizations are highly effective in practice.

\paragraph{Paper Outline}
The next section provides background about HNSW, disaggregated memory, and RDMA.
Section~\ref{sec:baseline} introduces our baseline and scalability constraints. 
In Sections~\ref{sec:optimizations-outline}-\ref{sec:query-routing}, we overview and discuss design choices to optimize our index such as node caching, logical index partitioning, and (adaptive) query routing.
Section~\ref{sec:evaluation} presents our experimental evaluation.
Finally, we review related work in Section~\ref{sec:related-work} and conclude the paper in Section~\ref{sec:conclusion}.

\section{Background}
\label{sec:background}

In this section, we provide the necessary background on the Hierarchical Navigable Small World index structure, the disaggregated memory architecture, and remote direct memory access.

\subsection{Hierarchical Navigable Small World}
Hierarchical Navigable Small World (HNSW)~\cite{malkov-20} is a graph-based index structure for approximate $k$-NN search that combines the ideas of navigable small worlds~\cite{malkov-14} and skip lists~\cite{pugh-90}.
Particularly, HNSW can be seen as an extension of skip lists, where each level represents a proximity graph rather than a linked list~\cite{malkov-20}.

\begin{figure}[t]
  \centering
  \includegraphics[width=0.8\linewidth]{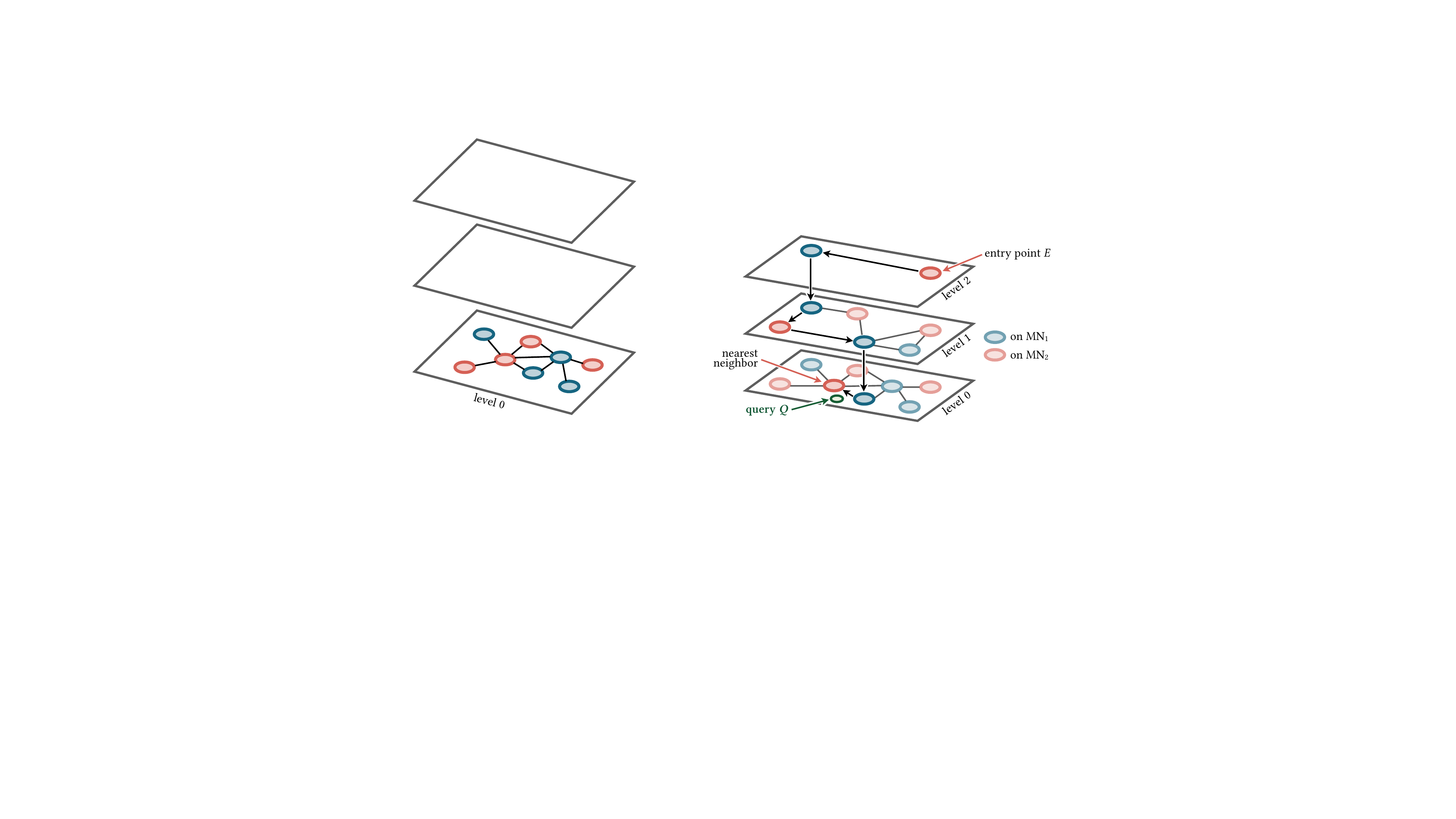}
  \caption{Concept of an HNSW index.}
  \label{fig:hnsw}
\end{figure}

\paragraph{HNSW Construction}
The index consists of $\log_M{(n)}$ levels, where $M$ is a construction parameter that determines the memory consumption and $n$ denotes the number of vectors in the dataset.
Upper levels consist only of a few nodes with long-range edges for quick navigation through the graph.
Contrastingly, the lowest level contains all nodes and short-range edges for fast localization of the $k$-nearest neighbors.
A node appears in level $l \geq 0$ with probability $1/M^l$ and vectors are inserted one by one during construction:
First, the level of a node is drawn from an exponentially decaying probability function.
Then, the node is inserted from top to bottom and connected to its $M$ closest neighbors w.r.t.\ the distance function.
Parameter $M$ guarantees the out-degree of a node, i.e., a node has at most $M$ outgoing neighbors per level.
Figure~\ref{fig:hnsw} illustrates the layered concept of HNSW.

\paragraph{HNSW Search}
Given a query vector $Q$, the search procedure always starts from a fixed entry point $E$ at the top-most layer.
The neighbors of $E$ (and their neighbors) are greedily traversed until the closest neighbor of $Q$ (i.e., $1$-NN) is identified.
The closest neighbor of level $l$ serves as an entry point for level $l-1$.
Note that if a node exists at level $l$ in the graph, it will also exist at all lower levels $[0, l-1]$.
This procedure is repeated until the bottom-most layer is reached, where all vectors are available and connected via short-range links that are used to identify the $k$-nearest neighbors of $Q$.

\subsection{Disaggregated Memory}
In conventional data centers, compute (CPU) and memory resources (RAM) are tightly coupled into monolithic hardware units.
Scaling compute power in response to workload peaks entails scaling main memory, resulting in over-provisioning.
This leads to underutilization of valuable resources~\cite{maruf-23,guo-19} and is highly undesirable as main memory is a key cost factors in today's data centers.

To overcome this problem, the disaggregated memory architecture physically decouples compute and memory into distinct hardware units that are connected via high-speed interconnects with support for remote direct memory access (RDMA) or Intel's Compute Express Link (CXL).
A compute node (CN) has many CPU cores but only small main memory, whereas a memory node (MN) is equipped with large amounts of main memory but near-zero compute power.
To achieve high scalability, computations are almost solely performed on CNs, which can directly access the main memory of MNs, e.g., using one-sided RDMA operations.

\subsection{Remote Direct Memory Access}
Remote direct memory access (RDMA) is a network communication mechanism that allows to directly access the memory of a remote machine without involving its CPU.
This enables very low access latencies (${\sim}2{\mu}s$) and is a key enabler for disaggregated memory.
RDMA operations are performed via so-called \emph{verbs} that can be either one-sided: \rdmaread, \rdmawrite, and atomic operations such as compare-and-swap (\rdmacas) and fetch-and-add (\rdmafaa); or two-sided: \rdmasend\ and \rdmareceive.
One-sided verbs require only one active side and do not involve the remote CPU.
Two-sided verbs are similar to socket-based communication primitives and involve the CPU of both sides.
That is, the remote machine must pre-post a \rdmareceive\ before another machine can issue a \rdmasend\ request.
Due to the near-zero compute power of MNs, most solutions for disaggregated memory exclusively rely on one-sided verbs.

\section{Baseline \& Limitations}
\label{sec:baseline}

This section introduces our baseline HNSW index design for the disaggregated memory architecture.
Importantly, we build a global HNSW index that preserves the original HNSW structure while being distributed to arbitrarily many MNs.
This allows us to achieve exactly the same accuracy, whereas other distributed approaches lose accuracy as they split the HNSW index.
First, we present the global index layout and describe how queries are processed, then we discuss limitations like scalability constraints and index updates.

\begin{figure}[h]
  \centering
  \includegraphics[width=\linewidth]{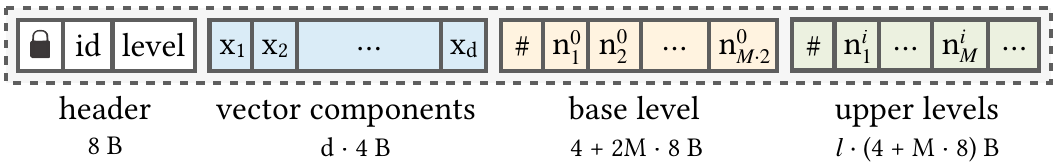}
  \caption{Physical memory layout of a single HNSW node.}
  \label{fig:layout}
\end{figure}

\subsection{Index Layout}
\label{ssec:index-layout}
A graph \emph{node} in HNSW stores the $d$-dimensional vector components and its outgoing neighbor nodes for each level on which the node occurs.
We represent them compactly in memory as shown in \figurename~\ref{fig:layout}.
To uniquely identify a node, we use a header that stores the node id, the maximum level at which this node appears, and a lock bit to lock the neighbor lists of this node for insertions.
Similar to the original HNSW implementation\footnote{\url{https://github.com/nmslib/hnswlib/}}, the lock is only used to construct the index concurrently.
For each level, we store a neighbor list prefixed by the number of list items.
A neighbor $n^i_j$ at layer $i$ is represented by a \unitbitadj{64} \emph{remote pointer:} The 48 least significant bits encode the virtual address of a memory region within the MN, and the remaining \unitbit{}s encode the identifier of the MN.
Hence, the remote pointer is a globally unique address among all MNs.
We leverage the fact that modern CPUs use only $2^{48}$~\unitbit{}s of the address space.

A node has at most $2M$ outgoing neighbors at the base level and $M$ neighbors on the upper levels, where $M$ is an HNSW construction parameter to control the memory consumption by constraining the out-degree.
As illustrated in \figurename~\ref{fig:layout}, the total size of a node is $8 + d \cdot 4 + 4 + 2M \cdot 8 + l \cdot (4 + M \cdot 8)$~Bytes , where $l \geq 0$ is the highest level at which the node occurs ($0$ if the node exists only at the base level).
When inserting a node, the node's maximum level $l$ is drawn from an exponentially decaying probability distribution.
Based on $l$, we allocate a fixed-size node on some MN.
For allocation, we randomly choose a MN on which we atomically increase a bump pointer (i.e., an \unitbyte{8} integer stored on a predefined address) using a remote \rdmafaa\ via RDMA.
To illustrate the concepts, we assume \unitbitadj{32} floating-point vector components (\unitbyte{4} in \figurename~\ref{fig:layout}), but our node layout generalizes to other encodings like \unitbitadj{64} floating points or integers.

To summarize, a node is stored on a randomly chosen MN, and its outgoing neighbors are represented by remote pointers (that point to other nodes) compactly stored in a fixed-size memory region.

\begin{algorithm}[t]
  \small
  \DontPrintSemicolon
  \KwIn{Query vector $q$, entry point $\mathit{ep}$, exploration factor $\mathit{ef}$, level $l$.}
  \KwOut{The $\mathit{ef}$-nearest neighbors to query $q$.}
  Insert $\mathit{ep}$ into next candidate set $C$ and top candidate set $T$\;

  \While{$C$ \upshape not empty\label{ln:term-1}}{
    $c \leftarrow$ pop closest vector from $C$; $t$ $\leftarrow$ get farthest vector from $T$\;
    \lIf{$d(q,c) > d(q,t)$\label{ln:term-2}}{break\tcp*[f]{all candidates evaluated}}
    \texttt{nlist\_rptr} $\leftarrow$ compute remote address from $c$ and $l$\; \label{ln:deduce-address}
    \texttt{neighborlist} $\leftarrow$ \texttt{RDMA\_READ(nlist\_rptr)} \label{ln:read-neighborlist}
    \vspace*{0.2em}

    \ForEach{\texttt{n\_rptr} $\in$ \texttt{neighborlist}\label{ln:for-each-neighbor}}{
      \lIf{\texttt{n\_rptr} $\in V$\label{ln:if-visited}}{continue}
      $V \leftarrow V \;\cup\;$ \texttt{n\_rptr}\;
      \texttt{neighbor} $\leftarrow$ \texttt{cache\_lookup(n\_rptr)}\; \label{ln:cache-lookup}
      \If{\texttt{neighbor} \upshape is \texttt{null}}{
        \texttt{neighbor} $\leftarrow$ \texttt{RDMA\_READ(n\_rptr)}\; \label{ln:read-neighbor}
        \lIf{\upshape admit to cache}{\texttt{cache\_insert(neighbor)}} \label{ln:cache-insert}
      }
      $t \leftarrow$ get farthest vector from $T$\; \label{ln:add-neighbors-begin}
      \If{$d(q,\texttt{neighbor}) < d(q,t)$ \upshape or $|T| < \mathit{ef}$\label{ln:dist-comp}}{
        $C \leftarrow C \;\cup\; \texttt{neighbor}$; $T \leftarrow T \;\cup\; \texttt{neighbor}$\;
        \lIf{$|T| > \mathit{ef}$\label{ln:add-neighbors-end}}{\upshape remove farthest vector from $T$}
      }
    }
  }
  \KwRet{$T$}

  \caption{Search Level}
  \label{algo:search-level}
\end{algorithm}

\subsection{Query Processing}
To process a query in HNSW, the nodes at the top-most level are greedily traversed (starting from a fixed entry point) until the nearest neighbor to a query $q$ is reached.
Typically, the entry point is the first node that has been inserted at the top-most level.
The nearest neighbor found at level $l$ serves as entry point for the next level $l-1$ as depicted in \figurename~\ref{fig:hnsw}.
Algorithm~\ref{algo:search-level} with $l=0$ shows the base level search, where the greedy search is expanded by searching the $\mathit{ef}$-nearest neighbors using two heap data structures based on the distance to the query: a min-heap $C$ to track the \emph{next candidates} and a max-heap $T$ to maintain the \emph{top candidates}.
The HNSW parameter $\mathit{ef}$ determines the search performance, where a higher $\mathit{ef}$ leads to more accurate results but slows down the search (note that $\mathit{ef} \geq k$).
The search routine greedily adds neighbors of candidates to $C$ and $T$ (lines~\ref{ln:for-each-neighbor},~\ref{ln:add-neighbors-begin}-\ref{ln:add-neighbors-end}) and terminates when no more candidates remain (line~\ref{ln:term-1}) or the farthest top candidate is closer to the query than the nearest next candidate (line~\ref{ln:term-2}).
Finally, the $k$-nearest neighbors are retrieved after returning $T$.
Note that Algorithm 1 is also used for the upper layers with $\mathit{ef}=1$ and $l>0$.
 
The only difference between \shine\ and standard HNSW search is that nodes and neighbor lists are read from remote memory (line~\ref{ln:read-neighborlist} and~\ref{ln:read-neighbor}).
The address of a neighbor list can be deduced from the node's remote pointer (line~\ref{ln:deduce-address}).

\subsection{Scalability Constraints}
In disaggregated memory, all computations are executed on the CNs. 
Thus, answering a single $k$-NN query involves fetching many high-dimensional vectors from the MNs to the CN that processes the query.
Note that the entire vectors are required for distance computations (line~\ref{ln:dist-comp} in Algorithm~\ref{algo:search-level}).
In our experiments, over \unit{6000} graph nodes with 200-dimensional vectors are traversed to process a single query of our largest dataset \tti.
As a result, the throughput is network-bound despite the use of an ultra-fast low-latency network.
The natural solution to this problem is to cache frequently accessed graph nodes on the CNs to reduce remote memory reads (lines \ref{ln:cache-lookup}-\ref{ln:cache-insert} in Algorithm~\ref{algo:search-level}).
The latency gap between local and remote memory is 10-100$\times$ lower than the gap between main memory and SSD.
Hence, an extremely fast cache with a low-overhead replacement strategy is crucial to reach high throughputs.

\begin{figure}[t]
  \centering
  \includegraphics[width=0.9\linewidth]{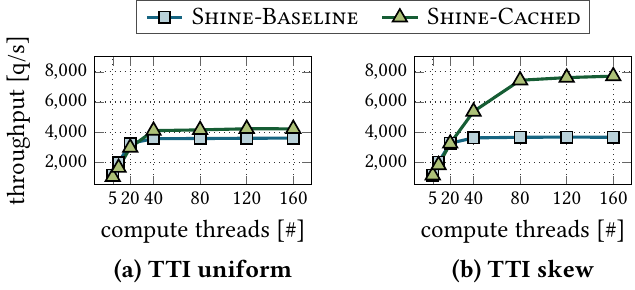}
  \caption{State-of-the-art caching for HNSW index traversals.}
  \label{fig:constraints}
\end{figure}

In contrast to many other index structures, an HNSW index in disaggregated memory poses a unique challenge to a caching mechanism: A single query requires access to many different graph nodes, in particular at the base level.
This results in many cache misses if (a) the cache is not sufficiently large (e.g., on CNs) or (b) the workload follows a uniform distribution.
In comparison, a B$^+$ tree point query merely requires $\log(n)$ node accesses. 
In \figurename~\ref{fig:constraints}, we plot the throughput (queries per second) over the number of compute threads for \shine\ with (\shinecached) and without (\shinebaseline) node caching\footnote{We use a state-of-the-art cache with cooling table~\cite{leis-18,lu-24}; cf.\ Section~\ref{sec:node-caching}.}.
We observe that adding a cache increases the throughput for skewed query workloads with a cache hit rate (CHR) of about 54\% (cf.\ \figurename~\ref{fig:constraints}b).
In contrast, the throughput for uniform query workloads is hardly improved with a CHR of merely 14\% (cf.\ \figurename~\ref{fig:constraints}a). 
See Section~\ref{sec:evaluation} for details on the experimental setup.
 
To address this problem and overcome the network bound, we propose to logically combine the caches of all CNs and route queries to the best matching CN (cf.\ Sections~\ref{sec:node-caching}--\ref{sec:query-routing}).

\subsection{Index Updates}
A limitation of HNSW lies in its static nature since the index has not been inherently designed for dynamic datasets.
The index structure allows updates, but particularly deletes and subsequent inserts are not very efficient and can lead to unreachable graph nodes~\cite{xiao-24}.
To alleviate this issue, it is common to rebuild the index after a certain amount of deletions~\cite{weaviate-vector-indexing}.
Supporting efficient index updates is considered an orthogonal research direction~\cite{xiao-24}, and we focus only on $k$-NN queries in this work.
Note that supporting insertions (without deletions) can be straightforwardly integrated into our method by using the insert procedure for constructing the graph, and locking neighbor lists during the search to guarantee consistent query processing.
Notably, our cache design (cf.\ Section~\ref{sec:node-caching}) is suitable for index updates and requires no coherency protocol (only neighbor lists are modified, which are not cached).

\smallskip
Next, we briefly overview optimization techniques for \shine\ and how to address the aforementioned scalability constraints.

\section{Optimizations Outline}
\label{sec:optimizations-outline}

We identify three key challenges \smallchallenge{1}{}--\smallchallenge{3}{} and briefly introduce three optimization techniques to address these challenges and optimize for scalability: (1) Node caching, (2) logical index partitioning, and (3) adaptive query routing.
For details, we refer to Sections~\ref{sec:node-caching}--\ref{sec:query-routing}.

\smallskip
\challenge{1}{Read Amplifications} 
HNSW requires to repeatedly fetch many high-dimensional vectors for distance computations, resulting in severe read amplifications.
\shine\ addresses this by caching frequently accessed nodes in the small memory of the CNs to save remote reads.

\smallskip
\challenge{2}{Cache Segmentation} 
The cache capacity of a CN is limited and shared among all workers on this CN.
In addition, the number of cache entries is restricted as high-dimensional vectors must be stored, and the caches of different CNs will largely overlap, thereby wasting valuable cache capacity and effectively reducing the overall cache size.
\shine\ logically combines the CN caches by assigning each CN to a specific portion of the index.
In particular, the HNSW graph is logically divided into $|\text{CN}|$ (number of CNs) many partitions, where each partition represents a CN's cache.
Once a query arrives at a CN, an \emph{oracle} determines the CN that can process the query most efficiently by finding the query's best fitting partition.

\smallskip
\challenge{3}{Query Routing} 
A query is processed by a random CN whose cache may not represent the target partition of this query.
Hence, a query must be forwarded to the CN with the best matching cache.
\shine\ implements adaptive query routing, where a query is forwarded to the CN returned by the oracle via a random MN using two-sided RDMA.
The destined CN receives the message and processes the query.
This strategy reduces the overlap of nodes among the caches, i.e., in total, more different nodes across the CNs are cached, which ultimately increases the hit rate and the throughput.

\begin{figure}[t]
  \centering
  \includegraphics[width=0.8\linewidth]{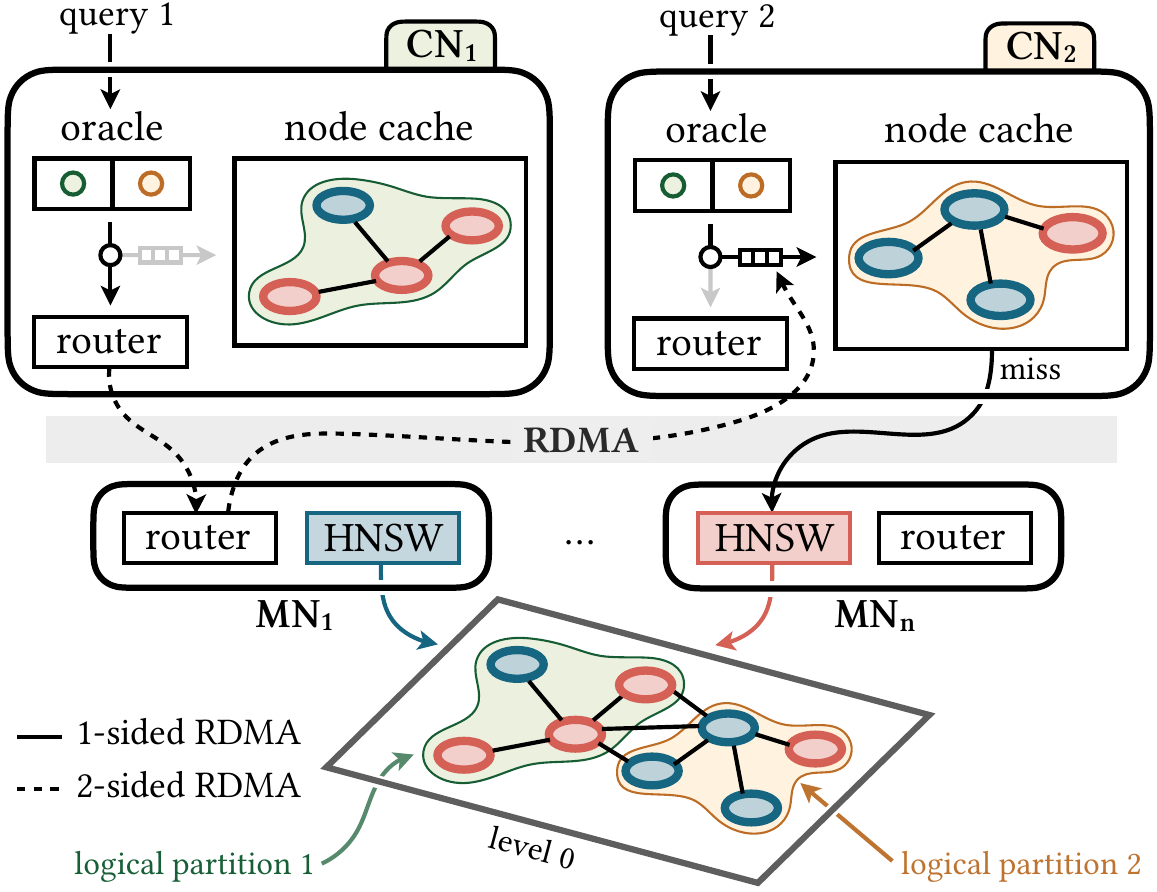}
  \caption{Overview of \shine\ and its optimizations.}
  \label{fig:overview}
\end{figure}

\begin{example}
  \figurename~\ref{fig:overview} shows the optimizations of \shine\ and exemplifies how queries are processed.
  Once CN$_i$ receives a query, \shine\ determines the best matching CN using the oracle.
  If the oracle returns $i$ (the local CN id; e.g., query~2 on CN$_2$), the query is enqueued to a local working queue, where it is further processed by CN$_i$'s compute threads.
  Otherwise (e.g., query~1 on CN$_1$), the \emph{router} sends the query and its destination (the best matching CN based on the oracle) to a random MN, which in turn sends the message to the desired CN. 
  Finally, the destined CN receives the message and enqueues the query into its local working queue.
\end{example}

\shine\ exploits the near-zero compute power of the MNs and performs routing solely with two-sided verbs, which avoids the need of a complicated synchronization mechanism (e.g., work stealing from other nodes).
Note that for scalability reasons, the CNs are only connected to MNs and not to other CNs in the disaggregated memory architecture.
Next, we discuss the cache design of \shine\ to reduce remote memory reads.
Logical index partitioning is presented in Section~\ref{sec:logical-index-partitioning} and adaptive query routing in Section~\ref{sec:query-routing}.

\section{Node Caching}
\label{sec:node-caching}

Processing a query involves reading numerous graph nodes and their neighbor lists from remote memory, resulting in heavily high read amplifications.
In particular, thousands of high-dimensional vectors are repeatedly fetched to determine the closest neighbors to the query by their distances.
This is expensive despite using a high-bandwidth, low-latency network, since accessing remote memory is still an order of magnitude slower than accessing local main memory~\cite{ziegler-22}.
To save remote memory accesses, \shine\ implements a lightweight software-level cache that keeps hot nodes in the local memory of a CN.

A major challenge is achieving efficient synchronization between cache accesses and replacements, which quickly becomes a bottleneck if the latency gap between local and remote memory is low.
We overcome this issue with lock-free reads and a lightweight replacement strategy.
Next, we discuss what to cache and how our cache is implemented.
Then, we elaborate on the choice of our cache replacement and admission policies.

\paragraph{What to Cache?}
In addition to caching vector components that are required for distance computations, neighbor lists (per level) can be cached.
However, since neighbor lists contribute only a small fraction of RDMA reads, we do not cache them.
For a node's neighbor list of size $M \cdot 8$~\unitbyte{}, we must read up to $M$ vectors, each of size $d \cdot 4$~\unitbyte{} (for simplicity, we omit the constant headers), where \unitbyte{8} is the size of a remote pointer and \unitbyte{4} the size of a vector component (e.g., a \unitbitadj{32} float) as shown in \figurename~\ref{fig:layout}.
Recall that $M$ is the HNSW construction parameter that determines the out-degree of a node and consequently, the space consumption of the index. 
Hence, the ratio of bytes read from remote memory for a neighbor list over the nodes is approximately $M \cdot 8 / (M \cdot ( d \cdot 4)) = 2 / d$.
For the base level, the ratio is the same but $M$ is replaced by $2M$ (cf.\ Section~\ref{ssec:index-layout}).
Thus, for 200-dimensional vectors, reading neighbor lists accounts only for 1\%.
In practice, this ratio is slightly larger because some nodes have already been visited and are not re-evaluated (cf.\ line~\ref{ln:if-visited} in Algorithm~\ref{algo:search-level}).

\paragraph{Cache Implementation}
The cache is implemented as a concurrent hash table mapping a remote pointer to a linked list of \emph{cache entries} to resolve hash collisions with chaining.
The list can be locked for insertions (i.e., admissions) and deletions (i.e., evictions), and a cache entry stores the key, a cooling flag (which is explained in Section~\ref{ssec:cache-replacement}), and a pointer to a local memory location that stores the vector.
Since cache accesses must be efficient due to the low latency gap between local and remote memory, cache lookups are lock-free, which is realized with pointer tagging:
A cache entry stores (1) a tag and (2) a tagged pointer (an \unitbyte{8} pointer, where the unused bits are used to encode a tag) to the subsequent entry in the chain.
The pointer tag must match the entry tag that the pointer references, otherwise the entry is corrupted (has been freed and potentially reused by another thread) and the cache lookup must be restarted.
Deleting a cache entry occurs under a lock on the list and increments the entry's tag, thereby invalidating the pointer that references that entry.
The released entry may then be reused.
We prefer pointer tagging over lock versioning as it invalidates individual cache entries instead the entire chain.

\subsection{Cache Replacement}
\label{ssec:cache-replacement}

Since the latency gap between local and remote memory accessed via RDMA (${\sim}2{\mu}s$) is notably smaller than the gap between main memory and SSDs (${\sim}10$-$100{\mu}s$), a cache replacement strategy with very low overhead is crucial.
Conventional replacement strategies, such as LRU, MRU, or frequency-based methods, must constantly update head and tail pointers of a centralized data structure (shared among all compute threads) to track information for future eviction. 
Moreover, in disaggregated memory, where the core count of CNs is typically high, maintaining such data structures, i.e., performing extra work for every cache hit, can lead to severe scalability bottlenecks.
For this reason, we choose a relaxed LRU variant following the design of~\cite{leis-18,lu-24} that randomly puts cache entries into a cooling state and avoids updating tracking information for each cache hit.
A cache entry remains in the cooling state for a certain amount of time and is eventually evicted if meanwhile no hit occurs.

The cooling mechanism is implemented via a hash table, where each bucket stores a fixed-size FIFO array, containing \unitbyte{8} keys (remote pointers) that represent entries in the cooling state, protected by a lock.
Rather than using a single cooling array, the hash table reduces the pressure of a single lock and thus increases scalability.
The size of the cooling table is relative to the cache size and keeps a fixed amount of cache entries (e.g., 10\%) in the cooling state.
 
\begin{figure}[t]
  \centering
  \includegraphics[width=0.9\linewidth]{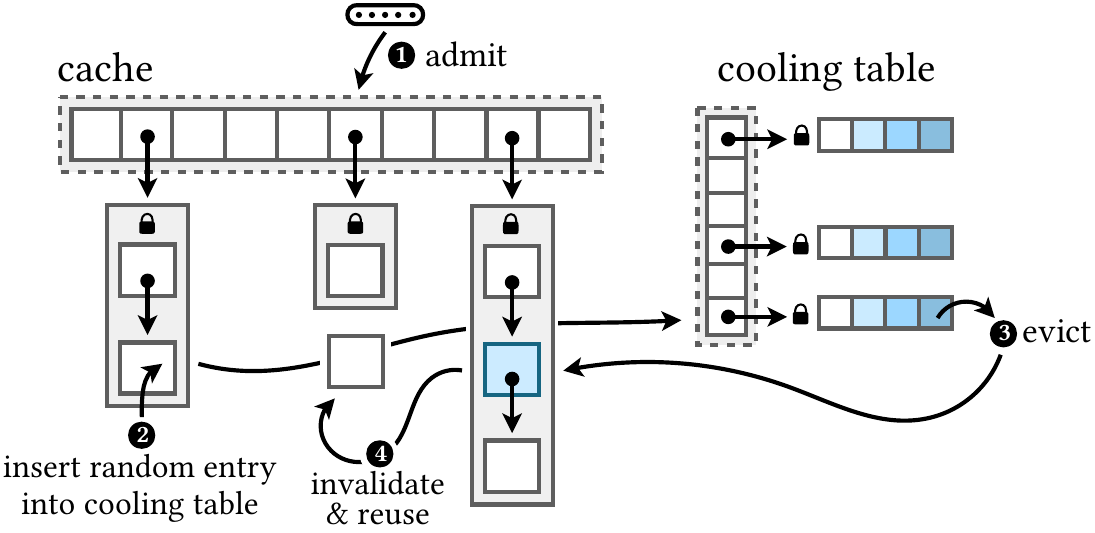}
  \caption{Cache implementation. Admitting a node to the cache triggers an eviction in the cooling table.}
  \label{fig:cache}
\end{figure}

\figurename~\ref{fig:cache} shows the functionality of the cache:
Once a node is admitted to the cache (we discuss cache admission in Section~\ref{ssec:cache-admission}) and no free cache entry is available (i.e., the cache is full), a cache entry is chosen at random and transitioned to the cooling state.
In particular, the cache entry is inserted into the cooling table by hashing its key and inserting the key at the beginning of the respective FIFO array.
If the array is full, the last entry drops out and is evicted.
For eviction, the hash of the key is computed to determine the bucket in the cache.
Finally, the bucket is locked and the cache entry is removed from the linked list.
Note that if multiple threads evict several entries concurrently, the evictions likely occur independently in different buckets, mitigating scalability bottlenecks of the cache data structure.
 
Hitting a cache entry that is currently cooling brings back the entry to the hot state.
To determine whether a cache entry's key is in the cooling table, we store an additional flag in the cache entry to indicate its cooling status.
On a cache hit, the key is removed from the cooling table and the flag is cleared.

\subsection{Cache Admission}
\label{ssec:cache-admission}

Since the amount of memory of CNs is small and caching nodes is inherently space consuming, it is important to keep only hot nodes in the cache.
If the cache is full, admitting a new node triggers an eviction of another node.
Many cache designs for disaggregated memory (e.g.,~\cite{wang-22,luo-23}) optimistically admit every item in the cache. 
But possibly, the evicted entry is hotter than the new item, which is unknown at time of admission.
Moreover, frequent evictions come with non-negligible synchronization overhead in the hash tables and thus reduce the overall cache performance.

To reduce the number of cache evictions, \shine\ admits base-level nodes only with a fixed probability as proposed by~\cite{lu-24}.
Our experiments have shown that admitting 1\% of the base-level nodes works well in practice, i.e., nodes that are accessed frequently will eventually make it to the cache.
Upper-level nodes that are crucial for navigation are always admitted.

\paragraph{Discussion}
Employing an efficient cache is inevitable to overcome the low latency gap between local and remote memory.
However, a cache is not enough to achieve high scalability for a HNSW-based index, because answering a single query requires many lower-level node traversals, leading to poor cache hit rates.
For instance, for a \emph{single} query on our \tti\ dataset, on average \unit{250} and \unit{5770} graph nodes are visited at the upper and at the base level, respectively.
In comparison, a B$^+$ tree point query requires at most $\log(n)$ node (or page) accesses, which is less than 5 (with a fanout of 64) for the dataset sizes we use.
Additionally, a single cache entry for a 200-dimensional floating-point vector requires \unitkb{0.8} of memory, significantly reducing the number of entries (i.e., vectors) that can be stored in the cache. 
This results in heavily high read amplifications: to answer a single query, almost \unitmb{5} must be fetched from remote memory, which is $1000\times$ more than reading $5\times1$~\unitkb{} pages to answer a B$^+$~tree point query using the same dataset size.
To improve the cache effectiveness, we logically combine the caches, which we discuss in the following sections.

\section{Dealing with Cache Segmentation}  
\label{sec:logical-index-partitioning}

This section describes the problem of cache segmentation in our distributed setting, which reduces the effective cache size, and proposes logical index partitioning as a solution.
The following discussion is based on a realistic model in which the queries are dispatched to CNs, such that each CN receives approximately the same number of queries, e.g., by assigning queries at random. 

As discussed in the previous section, caching for HNSW indexes is challenging due to the large number of nodes that must be cached. 
Moreover, the high dimensionality of the indexed vectors further constrains the number of cache entries that can be stored.
Therefore, it is crucial that the available cache memory is used effectively. 

The distributed nature of our setting makes it difficult to effectively utilize the globally available cache memory, i.e., the sum of all cache memories across all CNs. The cache memory of an individual CN is shared by all workers on that node, and the caches of the other CNs cannot be accessed. 
As all CNs process queries from the same global distribution, the overlap of cached nodes between different CNs is large, wasting valuable cache capacity.
Although the global cache size increases with more CNs, the effective cache size will not scale due to repetitive entries between caches on different CNs.

\subsection{Cache Segmentation Penalty}
\label{ssec:cache-segmentation-penalty}

We introduce a new concept, the \emph{cache segmentation penalty}, which quantifies the performance impact of dividing global cache memory among multiple nodes. While cache hit rate — defined as the number of cache hits divided by the total number of accessed nodes 
— is commonly used to assess cache performance, it fails to isolate the effect of segmentation: the hit rate is heavily influenced by the relative sizes of the cache and the index, making it difficult to attribute performance changes specifically to segmentation.

Intuitively, the cache segmentation penalty measures the ratio between the achieved cache hit rate (CHR) in a distributed system and the optimal hit rate (CHR$_\text{max}$) achievable in the same setting. 
The optimal rate corresponds to a scenario where no duplicate entries are stored across caches, modeled as a single cache shared by all CNs, with the same aggregated cache size over all CNs.

\begin{definition}[Cache Segmentation Penalty]
  Given $k=|\text{CN}|$ compute nodes with cache sizes $c_1, c_2, \dots, c_k$. Let CHR be the measured cache hit rate for a given query load in the distributed cache setting, and CHR$_\text{max}$ the cache hit rate achieved for the same setting except that all compute nodes share a single cache of size $C=\sum_{i=1}^{k}c_i$. The \emph{cache segmentation penalty} (CSP) is defined as:
  \begin{equation}
        \text{CSP} = 1-\frac{\text{CHR}}{\text{CHR}_\text{max}}  
  \end{equation}
\end{definition}

\begin{example}
  With 4 CNs, CHR$=25\%$ and CHR$_\text{max}=100\%$ indicate that all four caches store the same entries, yielding a CSP of 0.75, i.e., 75\% of the cache capacity is wasted. Similarly, CHR$=15\%$ and CHR$_\text{max}=60\%$ also result in a CSP$=0.75$, showing that the impact of cache segmentation is the same, despite a lower absolute hit rate.
\end{example}

\subsection{Logical Index Partitioning}

We propose logical index partitioning to mitigate the negative effects of cache segmentation.
The core idea is to divide the nodes of the HNSW index into disjoint partitions, assigning each CN to a unique partition. 
As a result, each CN becomes responsible for a specific portion of the index and primarily handles queries that are closest to its assigned partition. 
This specialization reduces overlap between caches, improving the cache segmentation penalty.

We propose a technique to efficiently cluster the index into partitions of similar size and discuss the \emph{oracle} used to assign queries to the most appropriate partition (i.e., CN). 
Note that the partitioning is purely logical: no data is moved, and the structure of the index (i.e., its nodes and edges) remains unchanged. 
The mechanism and policy of routing queries to CNs is discussed in Section~\ref{sec:query-routing}.

\paragraph{Clustering}
We use balanced $k$-means~\cite{maeyer-23,malinen-14} to cluster the nodes of the index into $k=|\text{CN}|$ (the number of CNs) many partitions of near-equal size.
Conventional $k$-means aims to build $k$ disjoint clusters such that the sum of squared distances of the data points to the cluster centroids is minimized.
Unfortunately, for high-dimensional vector spaces, often some very small clusters emerge, even if the data has a balanced distribution~\cite{bennett-00,maeyer-23}.
\emph{Balanced} $k$-means additionally minimizes the difference in cluster sizes and thus, ensures that each cluster consists of a similar number of points.
Since finding optimal clusters is NP-hard~\cite{aloise-09}, we use the approximation proposed by~\cite{maeyer-23} and cluster only on a small ($\leq$100k) subset of the nodes in the index.
In particular, we choose the first level of the HNSW index (from top to bottom) that contains $\geq$\qty{1000} nodes as a representative sample.

We found that for small and odd $k$ values (e.g., 3 or 5 CNs), the approximation may fail to generate well-balanced clusters.
To address this, we apply a simple yet effective heuristic: 
We first double the number of clusters until the resulting clusters are nearly equal in size.
Then, we greedily merge the pairwise closest clusters (based on centroid distance) until only $k=|\text{CN}|$ clusters remain.

Clustering is highly efficient (takes less than 1s) due to the small number of nodes involved. In our implementation, each CN performs clustering using the same random seed, ensuring identical cluster assignments. Only the cluster centroids are stored on each CN.

\paragraph{Oracle}
The oracle predicts the best fitting CNs for a given query $q$. It stores a mapping from the cluster centroids to the CNs and returns a list of CNs ranked by the distance between $q$ and the respective centroids. The CN that best fits $q$ is the first entry in the list, and the last entry is the CN with the most distant centroid. Depending on the routing policy, a query may not be assigned to the best fitting CN; therefore, the oracle maintains a ranked list.
Note that the prediction for a query requires only $k=|\text{CN}|$ distance computations.

Due to the low overhead of clustering, the oracle is updated with negligible overhead in response to changes in the number of CNs.

\section{Query Routing}
\label{sec:query-routing}

In order to improve the cache efficiency, \shine\ aims to process queries on the best matching CNs as proposed by the oracle  (cf.\ Section~\ref{sec:logical-index-partitioning}). In this section, we propose a lightweight routing mechanism and present different routing strategies.

The queries received by a CN (e.g., from clients) are appended to its \emph{input queue}. The CN pops queries from the input queue and either routes them to another CN or appends them to the CN's \emph{working queue} to be processed locally.
Since the disaggregated memory architecture does not allow direct communication between CNs (due to scalability reasons), MNs must be involved to route queries.

\subsection{Routing Mechanism}

In our disaggregated memory setting, we have two options to route queries through MNs: one-sided and two-sided RDMA verbs.

\paragraph{One-sided}
Index designs for disaggregated memory prefer one-sided RDMA verbs as two-sided verbs involve the CPU of the MNs (with low computation power). 
When using a one-sided approach, a routed query must be inserted into a queue associated to the target CN.
The queue is stored on a MN, requiring a total of $|\text{CN}|$ queues distributed across the MNs.
The queues must be synchronized (e.g., with a lock) because other CNs also enqueue queries; this requires at least an atomic operation (e.g., \rdmacaslong) and a subsequent \rdmawritelong.
The CN to which the query is routed, dequeues queries from its queue, which also requires synchronization.
Moreover, the memory locations of the queues must be known by all CNs, and new queues must be added if the number of CNs dynamically changes.

Overall, the synchronization effort required when using one-sided verbs is high relative to the lightweight task of routing. As RDMA atomic operations hardly scale under contention~\cite{kalia-16,ziegler-23}, \shine\ employs a two-sided approach.

\begin{figure}[t]
  \centering
  \includegraphics[width=0.975\linewidth]{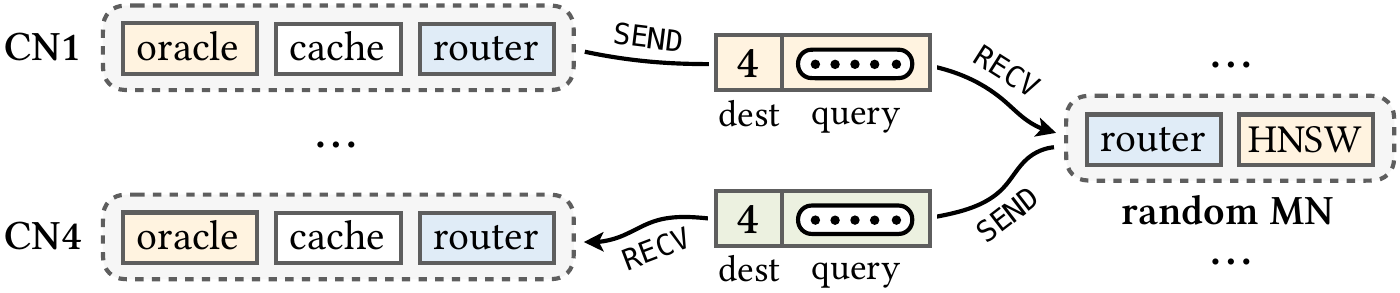}
  \caption{Example query routing from CN$_1$ to CN$_4$.}
  \label{fig:routing}
\end{figure}

\paragraph{Two-sided}
\shine\ leverages the low compute power of the MNs for routing:
Each CN and MN spawns a \emph{router} thread responsible for routing.
A CN's router wraps the query into a message and appends a header to encode the destination CN.
Then, the router \rdmasend s the message to a random MN; the MN is chosen randomly to evenly distribute the routing load.
Upon receiving the request, the MN reads the header and forwards the message (using a \rdmasend\ operation) to the target CN.
Finally, the destination CN receives the message and inserts the query into its local working queue. 
Figure~\ref{fig:routing} illustrates the routing process.

Despite their low compute power, MNs can handle the routing load in our setup. 
Recall that processing a single query with HNSW requires reading thousands of vectors from remote memory, resulting in significant read amplification. 
For example, answering a single query on one of our datasets (\spacev; 100-d) with 95\% recall requires reading approximately 1.6~\unitmb{} from remote memory (assuming no caching). 
Hence, a single MN with a \unitgbit{100} NIC can support a maximum throughput of about \unit{7800} queries per second, regardless of the number of CNs. 
The MN’s CPU only needs to route at least \unit{7800} queries per second, which is feasible given that routing involves merely reading the message header and forwarding the query — a lightweight task. 
A single core on our MNs can route about 167k queries per second.

\subsection{Routing Policies}

The overall goal of a routing policy is to increase the query throughput for a wide range of workloads. 

\begin{figure}[t]
  \centering
  \includegraphics[width=0.9\linewidth]{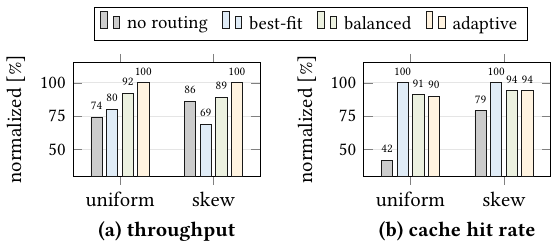}
  \caption{Comparison of query routing policies.}
  \label{fig:routing-comparison}
\end{figure}

A baseline that does not route any queries to other CNs suffers from a low cache hit rate, resulting in low throughput. This effect is particularly pronounced for uniform workloads (i.e., query loads that touch all parts of the index with similar probability) due to a high cache segmentation penalty. \figurename~\ref{fig:routing-comparison} exemplary shows the normalized throughput and cache hit rate for different routing strategies on the \spacev\ dataset with 5~CNs (cf.\ Section~\ref{sec:evaluation} for details on the experimental setup): Clearly, \emph{no routing} achieves the lowest relative hit rates for both uniform and skewed workloads. 

\paragraph{Best-Fit Routing}
The best-fit strategy relies on the oracle prediction for routing, where each query is routed to the best fitting CN. This approach significantly increases the cache hit rate, underpinning the effectiveness of our logical index partitioning technique and the oracle: For uniform workloads, best-fit routing increases the cache hit rate by a factor of almost $2.5\times$ w.r.t.\ a no-routing policy.

Unfortunately, the improved cache hit rate does not always translate into better throughput since skewed query workloads lead to execution skew, i.e., most queries are routed to a small subset of CNs responsible for the most frequently accessed partitions, creating bottlenecks. For the skewed workload in \figurename~\ref{fig:routing-comparison} (Zipf parameter $s=1$), the throughput of best-fit drops to 69\%, even below the throughput of the no-routing policy. 
This effect becomes even more pronounced with heavier skew: for Zipf parameters $s \geq 1.25$, the normalized throughput drops below 40\%.

\paragraph{Balanced Routing}
Best-fit routing significantly increases the cache hit rate but leads to unbalanced work packages that overload individual CNs. To avoid skewed loads, the \emph{balanced routing} policy assigns each CN the same number of queries. To this end, the queries in the input queue are routed in batches of size $b$.   

The router maintains a frequency counter that tracks the number of queries routed to every CN, including itself. 
Let $|\text{CN}|$ be the number of CNs, then each CN is sent at most $b/|\text{CN}|$ queries, leading to a balanced load of $b$ queries per CN and batch. The counter is reset at the end of each batch. In our experiments, we set $b=1000$.

The balanced routing policy aims to assign each query to its best matching CN, as determined by the oracle. If the top-ranked CN has already received its quota of $b/|\text{CN}|$ queries for the batch, the router continues down the oracle’s ranked list to find the next best CN that has not yet reached its limit.
As shown in \figurename~\ref{fig:routing-comparison}, the balanced approach reduces the normalized cache hit rate but improves the throughput from 80\% to 92\% for uniform queries and from 69\% to 89\% for skewed workloads.

\begin{algorithm}[t]
  \def\hist{H}
  \def\limits{L}
  \def\queue{Q}
  \def\progresses{P}
  \def\numcns{|\text{CN}|}
  \SetKwFunction{oracle}{oracle}
  \SetKwFunction{route}{route\_query}
  \SetKwFunction{sleep}{sleep}
  \newcommand{\lIfElse}[3]{\lIf{#1}{#2 \textbf{else}~#3}}

  \small
  \DontPrintSemicolon
  \KwIn{Batch size $b$, local CN id $c \in [0, \numcns)$, working queue $\queue$.}

  Initialize $\hist$ (histogram) and $\limits$ (limits) with arrays of size $\numcns$ and default values $0$ resp. $b/\numcns$. Initialize routed queries $r \leftarrow$ 0.

  \ForEach{\normalfont query $q$ received from clients}{
    \vspace*{0.25em}
    \tcp{determine closest CN that does not exceed the limit,}\vspace*{-0.2em}
    \tcp{where d is the distance to the centroid and i the CN id}
    $\delta \leftarrow \min \big\{ (d, i) \in \oracle(q) \;|\; \hist[i] < \limits[i] \big\}$\label{ln:oracle}\;
    \lIfElse{$\delta\; \texttt{!=}\; c$}{\route{$q, \delta, \text{MN}_\text{rand}$}}{$\queue\ \leftarrow \queue\ \cup q$}\label{ln:route-query}
    increment $\hist[\delta]$ and $r$ by 1\;

    \If(\tcp*[f]{batch processed}){$r\;\texttt{==}\;b$}{
      $p_\text{\tiny local} \leftarrow \queue\texttt{.size()}$\tcp*{track progress}\label{ln:inform-cns-begin}
      send $p\text{\tiny local}$ to all other CNs using random MNs\;\label{ln:inform-cns-end}
      \lWhile{$\queue\texttt{.size()} > t$}{\sleep{}\tcp*[f]{sync local threads}}\label{ln:sleep}

      $\progresses\ \leftarrow$ receive $p$ values from all CNs \tcp*{array of size $\numcns$}
      $\progresses[c] \leftarrow p_\text{\tiny local}$\;

      $\sigma \leftarrow \sum_{p \in \progresses}p$\tcp*{sum of progresses for normalization}\label{ln:update-limits-begin}
      \For(\tcp*[f]{update limits}){$i = 0$ \KwTo\normalfont $\numcns$}{
        $\omega \leftarrow \numcns\ \cdot (\sigma - \progresses[i])/\sum_{p \in \progresses} (\sigma - p)$\tcp*{weight}
        $\limits[i] \leftarrow \omega \cdot b/\numcns$\label{ln:update-limits-end}
      }
      reset histogram $\hist$ and set $r \leftarrow 0$\;
    }
  }
  \caption{Adaptive Query Routing}
  \label{algo:query-routing}
\end{algorithm}

\paragraph{Adaptive Routing}
Balanced routing assumes similar runtimes for all queries. This is not the case, though, due to differences in the cache performance: if a CN receives many residual queries, i.e., queries that are a poor fit according to the oracle, its cache performance degrades and leads to lower throughput.

To mitigate this problem, \emph{adaptive routing} dynamically adjusts the per-node query limit for each batch (fixed to $b/|\text{CN}|$ in balanced routing). Each CN periodically broadcasts the length $p$ of its local working queue to all other CNs. The relative queue lengths are then used to distribute the next batch of queries: CNs with longer queues receive fewer queries to balance the load.

\smallskip

Algorithm~\ref{algo:query-routing} outlines the adaptive query routing procedure. For each query, the oracle returns a ranked list of CNs, ordered by the distance between the query and the cluster centroids. The router selects the first CN $\delta$ in this list that has not yet reached its query limit for the current batch (line~\ref{ln:oracle}). If $\delta$ corresponds to the local CN, the query is added to its local working queue (line~\ref{ln:route-query}); otherwise, the query is forwarded to CN$_\delta$ via a random MN. 

After processing a batch of size $b$, the CN's router tracks the number of remaining queries $p$ in its local working queue and broadcasts $p$ (via a random MN) to all other CNs to inform them about its current progress (lines~\ref{ln:inform-cns-begin}-\ref{ln:inform-cns-end}).
Since the router thread on a CN only sends and receives queries, it progresses faster than the other threads that must answer queries.
Therefore, after tracking the progress value $p$, the router thread sleeps until the length of the local working queue falls below $t$ (line~\ref{ln:sleep}).
Note that the choice of $t$ is not critical as it only serves for synchronization. 
In our experiments, we set $t=1000$.

Finally, we update the limits for all CNs by computing a weight $\omega$ that captures the relative lengths of the working queues $p$. The weighted limits for each CN are computed as $\omega \cdot b/|\text{CN}|$ (lines~\ref{ln:update-limits-begin}-\ref{ln:update-limits-end}). Note that the weighted limits sum up to $b$, and  $\omega = 1$ indicates that all progress values $p$ are equal.

\smallskip
\figurename~\ref{fig:routing-comparison} shows that adaptive routing achieves the highest throughput for uniform and skewed query workloads, although the cache hit rate is slightly lower compared to balanced and best-fit routing.

In summary, query routing greatly improves the cache efficiency, which is particularly emphasized if the query workload follows a uniform distribution.
Moreover, adding adaptiveness to the routing mechanism significantly increases the query throughput.

\section{Evaluation}
\label{sec:evaluation}

In this section, we experimentally evaluate \shine\ and its optimizations.
First, we study the performance and scalability of \shine.
Specifically, the impact of caching and logically combined caches, i.e., logical index partitioning with adaptive query routing.
Finally, we perform a sensitivity analysis in which we vary several parameters such as the cache size ratio and the skewness of the query load.
The implementation of \shine\ is open source and publicly available\footnote{\url{https://github.com/DatabaseGroup/shine-hnsw-index}}.

\subsection{Experimental Setup}

\paragraph{Hardware}
We conduct all experiments on an 8-node cluster with 5~CNs and 3~MNs.
Each CN is equipped with two Intel Xeon E5-2630 v3 2.40~GHz CPUs with 16~physical (8~cores each) and 32~logical cores (hyperthreads), and \unitgb{96} of main memory.
The MNs have two Intel Xeon E5-2603 v4 1.70~GHz CPUs with 12~cores (6~cores each) and \unitgb{96} main memory.
All nodes run Debian~12 Bookworm on a Linux~6.1 kernel and are equipped with a Mellanox ConnectX-3 NIC connected to an 18-port SX6018/U1 InfiniBand switch (FDR \unitgbit{56}/s).
Since the NIC is installed on a single NUMA socket, we allocate memory on the socket of the NIC (using \texttt{numactl}\footnote{\url{https://linux.die.net/man/8/numactl}}) and assign compute threads on the CNs alternately to both CPU sockets to reduce NUMA effects.
Moreover, we use hugepages to reduce address translation cache misses on the NIC and share queue pairs to avoid queue pair thrashing~\cite{dragojevic-14}.
Similar to previous works~\cite{wang-22,luo-23,widmoser-24}, we model a disaggregated memory architecture by limiting the memory size of a CN to \unitgb{10} (shared among all compute threads) and assigning only a single CPU core to each MN.

\paragraph{Parameters}
For building the index, we set the construction parameters $\mathit{efC}=500$ and $M=32$.
For each dataset, we tune the search parameter $\mathit{efS}$ such that a recall of 95\% with $k=10$ nearest neighbors per query is reached, i.e., $R@10 \geq 0.95$.
Note that $\mathit{efS}$ is the candidate set size (cf.\ Algorithm~\ref{algo:search-level}).
Unless otherwise stated, we use a cache size ratio of 5\% (w.r.t.\ the index size) and spawn $5 \times 32=160$ compute threads across 5~CNs.
To mask the network latency, we use 4 coroutines per compute thread to yield CPU cores after issuing an RDMA operation.

\paragraph{Datasets}
We use five large-scale ANN datasets with 100 million vectors from the well-known BigANN benchmark~\cite{simhadri-21}, a public benchmark for which several companies (e.g., Microsoft, Meta) provide datasets from different domains, e.g., imaging, web search, and content-based retrieval.
We refer to \cite{simhadri-21} for a detailed description of the datasets.
\tablename~\ref{tab:datasets} compares the datasets w.r.t.\ the dimensionality (Dims), the distance function (L2 -- Euclidean; IP -- Inner product), the $\mathit{efS}$ parameter to reach $R@10 \geq 0.95$, and the index size accumulated over the MNs using our node layout (cf.\ \figurename~\ref{fig:layout}).

\begin{table}
  \centering
  \caption{Dataset characteristics.}
  \label{tab:datasets}
    \begin{tabular}{lccccr}
      \toprule
      \bf Dataset & \bf Dims & \bf Vectors & \bf Distance & \bf $\boldsymbol{\mathit{efS}}$ & \bf Index Size \\
      \midrule
      \bigann & 128 & 100~M & L2 & 80  & \unitgb{100} \\
      \deep   & 96  & 100~M & L2 & 100 & \unitgb{87} \\
      \spacev & 100 & 100~M & L2 & 100 & \unitgb{90} \\
      \tti    & 200 & 100~M & IP & 250 & \unitgb{126} \\
      \turing & 100 & 100~M & L2 & 150 & \unitgb{90} \\
      \bottomrule
    \end{tabular}
\end{table}

\paragraph{Queries}
\shine\ builds a global HNSW index that keeps all edges as in the original graph.
Hence, \shine\ achieves the recall of a single-machine HNSW index when constructed and run with the same parameters.
Consequently, our goal is not to evaluate the accuracy but rather the efficiency of our index.
Since only a few thousand queries are available for each dataset, we split 500k vectors from every dataset and use them as queries.
The query split is not part of the index and we use the same search parameter $\mathit{efS}$ that has been employed to reach a fixed recall of 95\% (cf.\ \tablename~\ref{tab:datasets}).
In addition to uniformly distributed queries, we generate skewed queries by following a Zipfean distribution (with default parameter $s=1.0$).
This allows us to evaluate the caching behavior when some queries occur more frequently than others.

For all runs, we use 100k queries to warm up the cache and the remaining 400k queries to benchmark the performance of the index.

\paragraph{Implementations}
In the following, we compare the performance of three \shine\ implementations with different optimization levels:
\newline\indent $\bullet$\,
\textbf{\textsc{\shinebaseline}:} Our \shine\ baseline implementation without optimizations like caching or adaptive routing (cf.\ Section~\ref{sec:baseline}).
\newline\indent $\bullet$\,
\textbf{\textsc{\shinecached}:} Our \shine\ implementation with caching enabled, i.e., one independent cache per CN (cf.\ Section~\ref{sec:node-caching}).
\newline\indent $\bullet$\,
\textbf{\textsc{\shinecachedaqr}:} Our final \shine\ implementation with all optimizations enabled, i.e., logical index partitioning (cf.\ Section~\ref{sec:logical-index-partitioning}) and adaptive query routing (cf.\ Section~\ref{sec:query-routing}).

\subsection{Performance and Scalability}

\begin{figure*}
  \centering
  \includegraphics[width=0.95\textwidth]{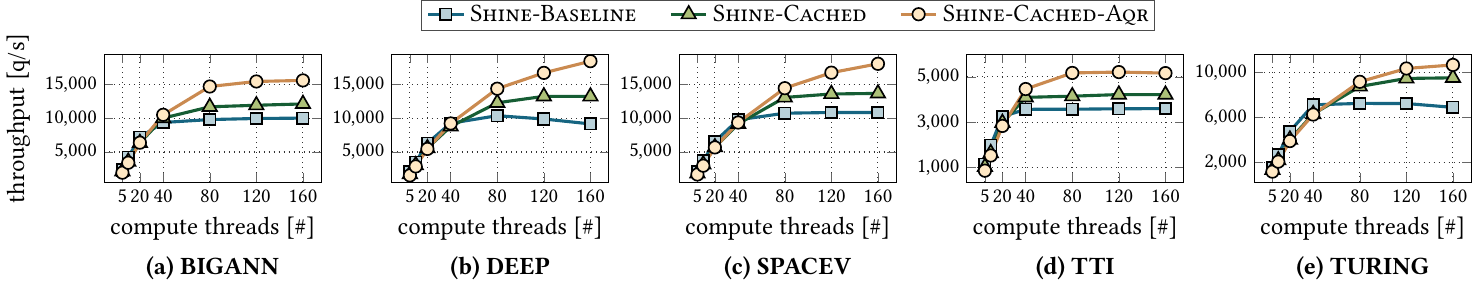}
  \caption{Throughput over increasing compute threads for uniform queries.}
  \label{fig:exp-scalability-uniform}
  \medskip
  \includegraphics[width=0.95\textwidth]{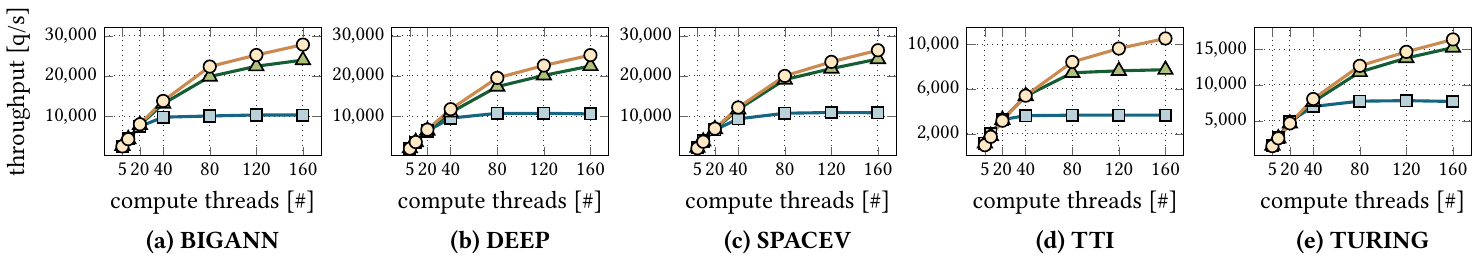}
  \caption{Throughput over increasing compute threads for skewed queries.}
  \label{fig:exp-scalability-skew}
\end{figure*}

In this set of experiments, we evaluate the throughput (in queries per second) over an increasing number of compute threads (5--160) for different query workloads and datasets.

\paragraph{Uniform Query Workload}
We use uniform query workloads to study the worst case for caching and to determine a lower bound on the performance of our proposed solution; \figurename~\ref{fig:exp-scalability-uniform} shows the corresponding throughput results.
For all datasets, we observe that our baseline implementation \shine\ is network-bound for $\geq\!80$ compute threads.
For \tti\ and \turing, the throughput of \shine\ already stagnates for $\geq\!40$ compute threads.
This is due to many high-dimensional vectors that must be fetched from the MNs over the network to find the $k$-NN at the base level.
\shine\ does not cache any vectors on the CNs.
Note that the absolute throughput for \tti\ compared to the other datasets is lower because the search exploration factor $\mathit{efS}$ is higher to reach the target recall of 95\% (cf.\ \tablename~\ref{tab:datasets}). 

For \shinecached, the throughput is slightly higher (on average by $1.3\times$) but still does not scale sufficiently well.
In general, a uniform query workload is the worst case for caching since nearly all base-level vectors (e.g., 90\% in \deep) in the HNSW graph are alternately visited with similar access frequencies.
The cache mainly improves the traversal performance for the upper levels, but has almost no positive impact for lower levels.

In contrast, \shinecachedaqr\ provides a significantly higher throughput for two reasons: (a) The caches of all 5~CNs are logically combined to enhance the overall cache size. (b) The queries are routed to the best matching CN.
For instance, \shinecachedaqr\ reaches a throughput of 18.1k~q/s with 160~compute threads on \spacev, which is an improvement of $1.7\times$ w.r.t.\ \shine\ and $1.3\times$ w.r.t.\ \shinecached, respectively.
Although \shinecachedaqr\ overcomes the network bound for most datasets, the uniform query workload is still inherently hard as no common base-level access pattern can be leveraged.
In this case, only employing larger cache size ratios allows to further improve the throughput (cf.\ Section~\ref{ssec:sensitivity-analysis}).

\paragraph{Skewed Query Workload}
\figurename~\ref{fig:exp-scalability-skew} shows the throughput for the realistic scenario of skewed query loads~\cite{lynch-88,walton-91}.
We observe that the baseline implementation \shine\ is network-bound and shows a similar behavior to uniform queries for the same reasons.
Adding a caching mechanism to \shine\ allows to break the network bound and significantly increases the throughput.
Frequently accessed vectors remain in the caches of the CNs due to the lightweight replacement strategy described in Section~\ref{sec:node-caching}.
For instance, on \bigann, the throughput of \shinecached\ is $2.3\times$ higher (from 10.3k~queries/s to 24k~queries/s).
Adaptive query routing of \shinecachedaqr\ further improves the throughput by 16\% to 27.8k~queries/s on \bigann.

\smallskip
In summary, adding caching is crucial to achieve high scalability for skewed workloads.
Logical index partitioning and adaptive query routing further increases the throughput by logically enhancing the overall cache size.
This is particularly important to overcome the network bound for workloads that tend to follow a uniform distribution.

\subsection{Impact of Caching with Adaptive Routing}
\label{ssec:cache-combination-efficiency}

\begin{table*}
  \def\cache{\textsc{Cached}}
  \def\routing{\textsc{Cached-Aqr}}
  \def\tp{q/s\;\;}
  \def\sep{0.55em}
  \centering
  \caption{Cache efficiency for different query loads. CHR denotes the cache hit rate and CSP the cache segmentation penalty.}
  \label{tab:csp}
    \begin{tabular}{llrrr@{\hskip \sep}|@{\hskip \sep}rrr@{\hskip \sep}|@{\hskip \sep}rrr@{\hskip \sep}|@{\hskip \sep}rrr@{\hskip \sep}|@{\hskip \sep}rrr}
      \toprule
      && \multicolumn{3}{c}{\bf\bigann} & \multicolumn{3}{c}{\bf\deep} & \multicolumn{3}{c}{\bf\spacev} & \multicolumn{3}{c}{\bf\tti} & \multicolumn{3}{c}{\bf\turing} \\ \midrule
       && CHR & CSP & \tp\ & CHR & CSP & \tp\ & CHR & CSP & \tp\ & CHR & CSP & \tp\ & CHR & CSP & \tp\ \\
      \midrule
      \multirow{2}{*}{\rotatebox[origin=c]{90}{unif}} & \cache\   & 18\% & 70\% & \SI{12130}{} & 21\% & 71\% & \SI{13241}{} & 21\% & 69\% & \SI{13692}{} & 14\% & 73\% & \SI{4219}{} & 18\% & 72\% & \SI{9540}{} \\
                                                      & \routing\ & 40\% & 33\% & \SI{15620}{} & 49\% & 32\% & \SI{18473}{} & 46\% & 35\% & \SI{18084}{} & 32\% & 38\% & \SI{5168}{} & 30\% & 55\% & \SI{10689}{} \\
      \midrule
      \multirow{2}{*}{\rotatebox[origin=c]{90}{skew}} & \cache\   & 61\% & 31\% & \SI{23950}{} & 63\% & 32\% & \SI{22480}{} & 69\% & 26\% & \SI{24220}{} & 54\% & 34\% & \SI{7715}{} & 60\% & 33\% & \SI{15208}{} \\
                                                      & \routing\ & 78\% & 12\% & \SI{27811}{} & 81\% & 12\% & \SI{25192}{} & 81\% & 12\% & \SI{26408}{} & 70\% & 14\% & \SI{10505}{} & 73\% & 18\% & \SI{16346}{} \\
      \bottomrule
    \end{tabular}
\end{table*}

In this section, we evaluate the efficiency of \shinecachedaqr, i.e., logical index partitioning combined with adaptive query routing, against the cache-only approach \shinecached\ for uniform and skewed query workloads.
In \tablename~\ref{tab:csp}, we show the cache hit rate (CHR), the cache segmentation penalty (CSP; cf.\ Section~\ref{ssec:cache-segmentation-penalty}), and the throughput (q/s) for each dataset.

For uniform query workloads, the CHR is generally low because most index nodes are recurringly accessed, leading to cache thrashing (i.e., continuous replacements).
Hence, only few cache entries (if any) can be reused.
For instance, on \deep, we merely reach a CHR of 21\% and a CSP of 71\%, meaning that 71\% of the overall cache capacity over all CNs are misspent for replicated entries.
Contrastingly, the CSP for \deep\ reduces to 32\% when logical index partitioning with adaptive query routing is enabled, increasing the throughput from 13.2k~q/s to 18.5k~q/s.
The performance gap is less pronounced for datasets with higher $\mathit{efS}$ values (e.g., \tti\ and \turing) because significantly more base-level nodes must be explored for a single query.
For example, the CSP on \turing\ reduces from 72\% to 55\%, but the throughput is only $1.12\times$ higher.

For skewed query workloads, the CHR is generally higher as many queries target similar parts of the HNSW index.
Thus, all CN caches are expected to hold reusable cache entries. 
Nonetheless, adaptive query routing decreases the CSP by 2-3$\times$ on all instances, resulting in higher throughput.
Note that the ideal CSP of 0\% is infeasible as most upper-level nodes are stored in all caches for quick navigation.

In summary, logical index partitioning with adaptive query routing consistently reduces the CSP and provides higher throughput.

\begin{figure}
  \centering
  \includegraphics[width=0.95\linewidth]{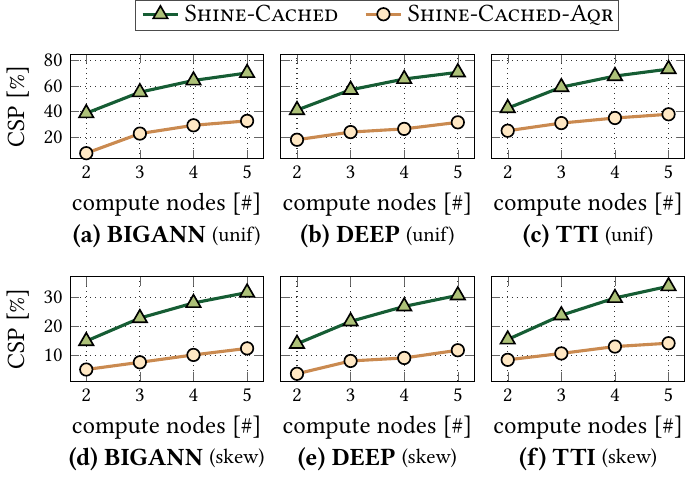}
  \caption{Cache segmentation penalty (CSP) over increasing compute nodes for uniform and skewed query loads.}
  \label{fig:exp-csp}
\end{figure}

\paragraph{More Compute Nodes}
Next, we study the CSP of \shinecached\ and \shinecachedaqr\ over an increasing number of physical CNs.
Recall that our goal is to logically combine the caches of the CNs to increase the number of cached entries and to improve the overall cache efficiency.
In \figurename~\ref{fig:exp-csp}, we observe that the behavior between uniform (cf.\ \figurename~\ref{fig:exp-csp}a-c) and skewed workloads (cf.\ \figurename~\ref{fig:exp-csp}d-e) is similar.
Over all instances, the cache efficiency of \shinecachedaqr\ is on average $2.3\times$ higher compared to \shinecached. 
The CSP increases faster for \shinecached\ because the cache hit rate does not change with increasing CNs.
In contrast, for \shinecachedaqr, the cache hit rate increases but grows slower than the overall cache capacity when adding CNs.
Ideally, the CSP would stagnate (or even decrease if all accessed nodes fit into the combined caches).
We attribute the CSP increase to the fact that adaptive routing not only optimizes the query placement, but also balances the execution load.
Moreover, finding an optimal (logical) graph partitioning is inherently hard.
Nonetheless, \shinecachedaqr\ efficiently leverages the overall cache capacity of all CNs and considerably reduces the CSP.

\subsection{Sensitivity Analysis}
\label{ssec:sensitivity-analysis}

In the following experiments, we evaluate the performance impact of (a) varying cache size and (b) varying skewness.

\begin{figure*}
  \centering
  \includegraphics[width=0.95\textwidth]{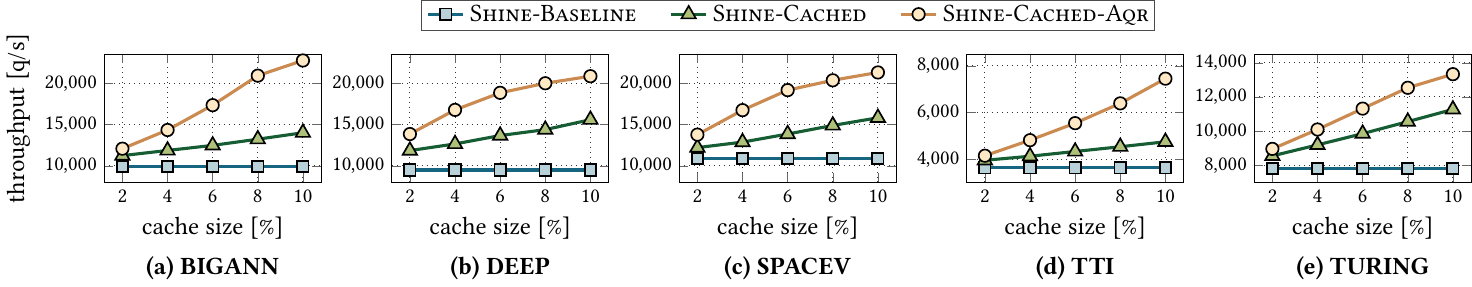}
  \caption{Impact of cache size on query performance for a uniform workload.}
  \label{fig:exp-cache-size}
  \medskip
  \includegraphics[width=0.95\textwidth]{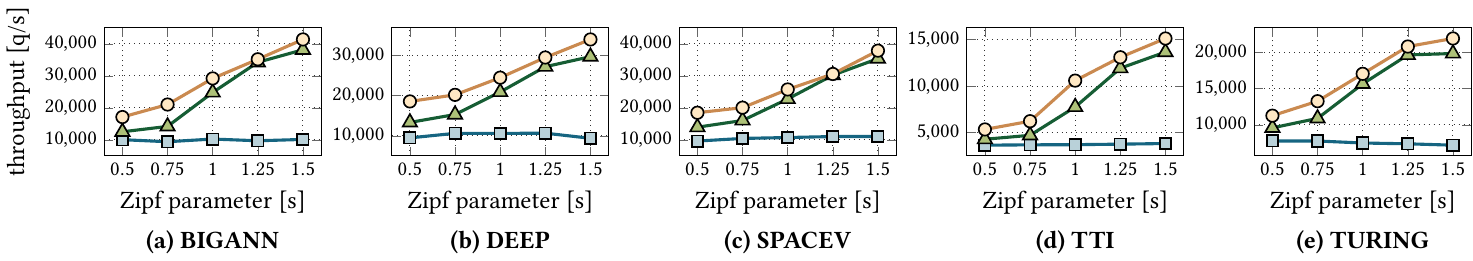}
  \caption{Impact of skew on query performance.}
  \label{fig:exp-skew}
\end{figure*}

\paragraph{Cache Size Ratio}
\figurename~\ref{fig:exp-cache-size} shows how the cache size ratio (i.e., $\frac{cache\ size}{index\ size}$) influences the throughput.
We start with a 2\% cache and increase the ratio up to 10\%.
For instance, on \deep, a 10\% cache size ratio translates into a \unitgb{8.7} cache per CN (cf.\ \tablename~\ref{tab:datasets}).
As a reference, we show the baseline implementation of \shine\ (without cache).
We study the cache performance for uniform queries to emphasize the positive impact of logical index partitioning with adaptive query routing (which is less significant under skew, cf.\ \figurename~\ref{fig:exp-scalability-skew}).
Certainly, a larger cache results in more cached HNSW nodes on the CNs and higher throughput.
Notably, adaptive query routing further boosts the throughput significantly.
For \bigann, employing a 10\% cache increases the throughput by $1.4\times$, whereas \shinecachedaqr\ improves the throughput by $2.3\times$ over the baseline.

\paragraph{Skewness}
We study the performance impact for increasingly skewed workloads.
\figurename~\ref{fig:exp-skew} shows the throughput over an increasing Zipf parameter $s \in \{0.5, 0.75, 1.0, 1.25, 1.5\}$ for each dataset.
The throughput of our baseline implementation \shine\ remains almost unchanged, as skew primarily influences the cache performance.
\shinecached\ and \shinecachedaqr\ perform significantly better under skew because many cached nodes are reused over subsequent queries, leading to fewer cache misses. 
The performance under high skew is almost on par since most required nodes fit into the cache of a single CN.
A very high skew basically compensates the (positive) effect of adaptive query routing.
For instance, on \spacev\ with $s \geq 1.25$, the CHR of \shinecached\ is already $\geq 94\%$.
Moreover, we clearly observe that adaptive query routing is a lightweight task (cf.\ Section~\ref{sec:query-routing}) and does not harm the performance.

\smallskip
In summary, caching is inevitable to improve the performance for skewed query workloads.
Logical index partitioning and adaptive query routing significantly increase the throughput on all instances, particularly if the cache performance of a single CN is low.

\subsection{Discussion}

As confirmed in our evaluation, caching on CNs is crucial to achieve high scalability and deal with network bandwidth limitations.
However, caching HNSW nodes is challenging because high-dimensional floating-point vectors are inherently space consuming, while the main memory of the CNs is strictly constrained in disaggregated memory.
Moreover, many different nodes, particularly at the base level, must be accessed for answering a single query.
For instance, more than \unit{6000} nodes are visited to process a single query for \tti.
In comparison, a B$^+$ tree point query merely requires $\log(n)$ node accesses, which is less than 5 (with a fanout of 64) for our dataset sizes.

\paragraph{Cache Compression}
To further increase the cache capacity, a natural direction might be to compress the nodes stored in the cache by employing a lossless compression scheme.
However, floating-point vectors (that are used for nearest neighbor search) are often already represented tightly or even have been compressed with lossy compression techniques such as PCA (e.g., \deep).
For this reason, only very low compression ratios can be achieved such that trading compute power for decompression does not pay off.
Although many compression algorithms provide extremely fast decoders~\cite{chen-24}, the cost of accessing nodes from remote memory via RDMA is lower, implying that cache compression can only be advantageous if the compression ratio is sufficiently high.
Note that compressing all nodes in the index on the MNs is orthogonal to our approach and trades compute power (decoding overhead) for more network bandwidth.

\section{Related Work}
\label{sec:related-work}

In this section, we review two related topics: scalable ANN search and index structures for disaggregated memory.

\paragraph{Scalable ANN}
Distributed approaches first partition large datasets into multiple \emph{shards} until each shard fits into main memory. 
Then, they send a query to (a subset of) relevant shards for processing, and finally combine the results.
Deng et al.~\cite{deng-19} propose Pyramid, a distributed HNSW-based solution that builds a small meta-HNSW to capture the structure of the entire dataset, which is used to identify shards likely containing the neighbors of a query.
On each shard, a local HNSW index is employed.
Liu et al.~\cite{liu-25} employ the idea of Pyramid in a disaggregated memory scenario. 
The graph partitions are stored on the memory nodes, while the meta-HNSW is cached on the compute nodes.
The most relevant shards w.r.t.\ the meta-HNSW are fetched to the compute node that handles the query.
In addition, queries are processed in batches to reduce network bandwidth usage by avoiding multiple transfers of the same partitions when queries in a batch have overlapping shards.
Doshi et al.~\cite{doshi-21} propose LANNS, a Spark-based method that uses a learned hyperplane to split the data into shards and also utilizes local HNSW indexes.
Muja~\& Lowe~\cite{muja-14} propose FLANN that uses random sharding and a $k$-means tree as an in-memory index per shard.
Since all methods partition the graph structure and potentially remove important edges between components, they are less accurate than the original HNSW index.
In contrast, \shine\ builds a global HNSW graph that keeps all edges, and thus reaches the same accuracy as a single-machine HNSW index.

Jang et al.~\cite{jang-23} propose CXL-ANNS, an emulated disaggregated graph-based approach that enhances host memory using Compute Express Link (CXL).
CXL-ANNS statically caches nodes close to the entry point in local memory to minimize the latency of far-memory accesses.
The authors prototype the CXL memory pool on FPGAs as currently no commercially available fully functional CXL system exists.
Ko et al.~\cite{ko-25} propose COSMOS, an ANNS system that integrates programmable cores and rank-level processing units within CXL devices to maximize memory bandwidth and reduce PCIe traffic.
Performance evaluations are simulated as CXL lacks commodity products on the market.
In contrast, \shine\ runs on commodity RDMA NICs.

Hierarchical disk-based approaches \cite{subramanya-19,ren-20,gollapudi-23,chen-21,wang-24,tian-24} store the index and vectors on (Smart)SSDs and keep a simplified (or compressed) index for navigation in main memory.
First, a set of candidates is determined by using the in-memory index structure. 
Then, the result is refined by re-examining the candidate set and accessing the actual vectors on disk.
However, for very large datasets, such approaches lack accuracy (due to their simplified/compressed main-memory index) and fail to scale computationally as they operate on a single machine.
\shine\ retains the entire index structure in (remote) main memory, preserving accuracy and enabling independent scaling of memory and compute resources.

Recent studies implement ANN search using GPUs~\cite{johnson-21,groh-23,yu-22,zhao-20} and FPGAs~\cite{jiang-23,zeng-23} to exploit their highly parallel computing capabilities.
However, due to their limited memory capacities, these solutions generally fail to scale to very large datasets.
\shine\ is designed for the disaggregated memory architecture, and thus independently scales to both memory and compute resources.

\paragraph{Index Structures for Disaggregated Memory}
In recent years, numerous index structures for disaggregated memory have been proposed including B$^+$ trees~\cite{lu-24,ziegler-19,wang-22,an-23,wang-25}, radix trees~\cite{luo-23}, LSM-trees~\cite{wang-23}, learned indexes~\cite{li-23}, hybrid indexes~\cite{luo-24}, hash tables~\cite{zuo-21}, and inverted list indexes~\cite{widmoser-24}.
Many index structures implement a software-level cache to reduce latency.
However, caching in \shine\ is more challenging because a single query requires traversing significantly more nodes (particularly at the base level) compared to most other index structures.
Moreover, \shine\ is the first graph-preserving HNSW index structure for disaggregated memory.

\section{Conclusion}
\label{sec:conclusion}

In this work, we proposed a scalable HNSW index for ANN search in disaggregated memory.
Unlike existing distributed approaches that partition the graph structure at the cost of accuracy, our solution builds a global HNSW index over arbitrarily many memory nodes that preserves the original HNSW graph.
Therefore, we are able to achieve the same accuracy as a single-machine HNSW.
Answering HNSW queries is challenging due to severe remote read amplifications that drastically constrain query processing performance.
To reduce remote memory reads, we cache hot nodes on the compute nodes.
However, HNSW access patterns demand a large number of nodes that must be cached, effectively rendering conventional caching impractical for uniform workloads. 
Therefore, we proposed to logically combine the caches of the compute nodes to increase the overall cache capacity and achieve high scalability.
Our experimental evaluation empirically demonstrated the scalability and efficiency of our solution for numerous datasets with changing workloads.


\balance
\bibliographystyle{ACM-Reference-Format}
\bibliography{references}

\end{document}